\documentclass[aps,prl,reprint,superscriptaddress,showpacs]{revtex4-1}

\usepackage{amsmath,amsfonts,bm,graphicx,verbatim,mathrsfs,units}

\usepackage{color,colordvi}
\definecolor{blue}{rgb}{0,0,1}
\definecolor{grey}{rgb}{0.6,0.6,0.6}

\begin{document}

\title{Negative Full Counting Statistics Arise From Interference Effects}

\author{Patrick P. Hofer}
 \affiliation{Department of Physics, McGill University, Montreal, Quebec, Canada H3A 2T8}
 \affiliation{D\'epartement de Physique Th\'eorique, Universit\'e de Gen\`eve, 1211 Gen\`eve,
  Switzerland}
  
\author{A. A. Clerk}
 \affiliation{Department of Physics, McGill University, Montreal, Quebec, Canada H3A 2T8}

\date{\today}

\begin{abstract}
The Keldysh-ordered full counting statistics is a quasi-probability distribution describing the fluctuations of a time-integrated quantum observable.  While it is well known that this distribution can fail to be positive, the interpretation and origin of this negativity has been somewhat unclear.  Here, we show how the full counting statistics can be tied to trajectories through Hilbert space, and how this directly connects negative quasi-probabilities to an unusual interference effect.  
Our findings are illustrated with the example of energy fluctuations in a driven bosonic resonator; we discuss how negative quasi-probability here could be detected experimentally using
superconducting microwave circuits.
\end{abstract}

\pacs{42.50.Lc, 72.70.+m, 03.65.Ta}

\maketitle

\textit{Introduction} -- Quasi-probability distributions such as the Wigner function \cite{wigner:1932} are powerful tools that allow one to visualize quantum states in phase space. They have played a seminal role in quantum mechanics since the early beginnings of the theory. Among their many uses are the identification of non-classical states: these are states where the Wigner function
(or some other related distribution) fails to be positive definite (see, e.g.~\cite{qm_phase:book}). Such non-classicality can constitute a resource for quantum information processing \cite{howard:2014, veitch:2012}. 

Recently, a very different kind of quasi-probability distribution has found widespread utility, the so-called full counting statistics (FCS) \cite{levitov:1996,nazarov:book2,nazarov:2003}. Unlike the Wigner function, the FCS does not describe the instantaneous state of a quantum system, but rather describes its time-history and dynamics: it characterizes the fluctuations of a time-integrated quantum observable.
As has been discussed extensively, the FCS distribution describes the ``intrinsic" fluctuations of the system absent any coupling to a measurement device \cite{nazarov:2003,clerk:2011}.  Nonetheless, it can be used to directly predict the outcome of realistic measurement setups, where the added noise of the measurement combines with the intrinsic system fluctuations to determine the final measured distribution \cite{nazarov:2003,bednorz:2010,clerk:2011,bednorz:2012}.
FCS first arose in the study of current fluctuations in quantum electronic conductors, where the transmitted charge is the time-integral of the current operator  \cite{levitov:1996,nazarov:book,nazarov:book2}; it continues to be a crucial tool in quantum transport, and has also been used to characterize cold atom systems \cite{gritsev:2006}, work \cite{solinas:2015} and heat fluctuations \cite{saito:2007}, dynamical phase transitions of classical systems \cite{flindt:2013,flindt:2014}, and quantum-optical systems \cite{clerk:2011}.  FCS have also recently been connected to weak measurement theory \cite{wei:2008,bednorz:2010}. 


Similar to conventional quasi-probability distributions, the FCS distribution can fail to be positive-definite. As the FCS describes the time-history of a system, negativity here is indicative of the presence of non-classical temporal correlations and/or dynamics which render a backaction free measurement impossible \cite{bednorz:2010,clerk:2011}.
Largely because many of the most studied systems are immune to backaction (e.g. gauge invariant electronic transport at long times), and thus described by a positive definite FCS  \cite{nazarov:2003}, very little work has been undertaken on the meaning, origin or utility of negative FCS; notable exceptions are  \cite{belzig:2001,bednorz:2010,belzig:2010,clerk:2010,clerk:2011,bednorz:2012}. Considering the utility of negativities in more conventional quasi-probabilities, it is desirable to obtain a better understanding of negative FCS.

\begin{figure*}[t!]
\centering
\includegraphics[width=\textwidth]{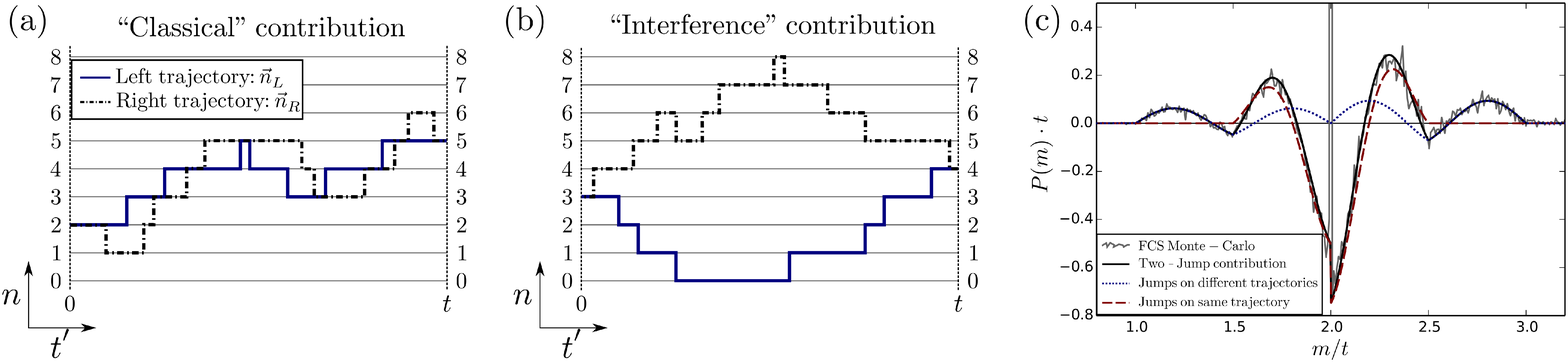}
\caption{FCS for a bosonic resonator and contributing pairs of trajectories. (a) Illustration of a pair of trajectories contributing to $P(m_0)$, where the $m$-value of each trajectory is the same: $m_L=m_R=m_0$. Such pairs yield a positive contribution. (b)  Illustration of a pair of trajectories with $m_L\neq m_R$ but $(m_L+m_R)/2=m_0$. As discussed in the text, such a pair can yield a negative contribution to $P(m_0)$.
(c) FCS for a cavity initially prepared in the $n_0=2$ Fock state. The analytical result (black, solid) consists of a contribution where the jumps are located on different trajectories (blue, dotted) and a contribution where the jumps are located on the same trajectory (red, dashed). The singular, zero jump contribution of Eq.~\eqref{eq:pofm0} is omitted. A Monte-Carlo simulation (grey) using 50,000 trajectories is in good agreement with the analytical results. Parameters: time $t=4/\Delta$, drive strength $f=\Delta/16$, where $\Delta$ is the drive detuning.}
  \label{fig:trajscem}
\end{figure*}


In this work we present a clear physical picture for how negative FCS emerge. We connect the FCS distribution to trajectories the system takes through Hilbert space. The resulting expression gives an intuitive understanding of the microscopic processes which contribute to the FCS and allows us to show that negative FCS are the direct result of an unusual interference phenomena: the interference of amplitudes associated with two trajectories can contribute to the quasi-probability, even though the classical probabilities for each trajectory do not contribute.  Our approach also demonstrates why negative FCS in general requires systems where a few degrees of freedom are relatively isolated. We stress that in contrast to Refs.~\cite{nazarov:2003,lorenzo:2004,lorenzo:2006,bednorz:2008},
our main focus is to understand the negativity in the FCS, and not on how the inclusion of detectors modifies the FCS and restores positivity in the final measured distribution. Nonetheless, our approach also gives an intuitive picture of this process (see supplemental material (SM) \cite{supplementarx}). Our approach is particularly well-suited to investigating the short-time FCS, a regime which is relevant to fast experimental protocols but that has received only limited attention.

To make the utility of our approach clear, we focus on a particularly simple system that exhibits negative FCS:  the time-integrated energy fluctuations in a coherently driven bosonic single-mode resonator.  The FCS here are particularly amenable to experimental measurement, and their negativity was recently discussed as a potentially powerful way to detect non-classical behavior in an optomechanical system \cite{clerk:2010}. Finally, we analyze a realistic circuit quantum electrodynamics (cQED) measurement setup for detecting negative FCS.

\textit{Definition of FCS} -- We consider an observable $\hat{n}(t)$ in the Heisenberg-picture, and are interested in characterizing the fluctuations of its time integral $\hat{m}=\int\limits_{0}^{t}dt'\hat{n}(t')$. Since $\hat{n}(t)$ does not necessarily commute with itself at different times, the higher moments of $\hat{m}$  will be contingent on 
how one chooses to time-order the various factors of $\hat{n}$.  The well developed field of FCS resolves this ambiguity by considering how 
one would measure $\hat{m}$; guided by this, the appropriate moment generating function for $m$ is \cite{levitov:1996,nazarov:2003,nazarov:book2} ($\hbar=1$)
\begin{equation}
\label{eq:momgen}
	\Lambda(\lambda)\equiv\int dm P(m)e^{-i\lambda m}\equiv{\rm Tr}\left\{e^{-i\hat{H}_\lambda t}\hat{\rho} e^{i\hat{H}_{-\lambda} t}\right\},
\end{equation}
where $P(m)$ is the quasi-probability distribution of interest (the FCS), $\hat{\rho}$ is the system density matrix at $t=0$, and $\hat{H}_\lambda=\hat{H}+\lambda\hat{n}/2$ with $\hat{H}$ being the Hamiltonian of the system. 

A simple way to motivate Eq.~(\ref{eq:momgen}) is to consider an idealized measurement  where an auxiliary qubit couples to $\hat{n}$ via 
$\hat{H}_c=\lambda\hat{n}\hat{\sigma}_z/2$ \cite{levitov:1996}.  If $\hat{n}$ were a classical stochastic variable $n(t)$, the qubit would precess by an angle $\lambda m=\lambda\int\limits_{0}^{t}dt'n(t')$, and the off-diagonal reduced density matrix element would directly yield the average of $\exp(-i\lambda m)$, i.e.~the moment generating function.  This then motivates Eq.~\eqref{eq:momgen} in the quantum case.  
 This is only one of several idealized measurement schemes which lead to Eq.~\eqref{eq:momgen} \cite{nazarov:2003,clerk:2010,clerk:2011}. 
Eq.~(\ref{eq:momgen}) can also be motivated by the Keldysh path integral approach \cite{kamenev:book}. The time-ordering of $\hat{n}(t)$, which ultimately leads to the negativities in the FCS, is thus dictated by the fact that the FCS is a measurement-independent quantity.

\textit{Unravelling the FCS} -- In the spirit of Feynman's path integral approach, we now divide the time evolution in Eq.~\eqref{eq:momgen} into $N$ infinitesimal steps of duration $\delta t$; between these partitions, we introduce resolved identity operators.
We start with the simplest case, where $\hat{n}$ has a discrete spectrum and further, where our system has no additional quantum numbers, such that $\mathbb{I}=\sum_n|n\rangle\langle n |$ is the identity operator. 
Inserting the identities allows us to replace the operator $\hat{n}$ by its eigenvalues. The FCS can then be obtained by Fourier transforming Eq.~\eqref{eq:momgen}
\begin{equation}
\label{eq:pofmres}
\begin{aligned}
P(m)=\sum_{\vec{n}_L,\vec{n}_R}&\delta_{n_f^L,n_f^R}\delta\left(m-\frac{1}{2}m_L-\frac{1}{2}m_R\right)\\&\times\langle n_1^L|\hat{\rho}|n_1^R\rangle A(\vec{n}_L)A^*(\vec{n}_R),
\end{aligned}
\end{equation}
with the amplitudes
\begin{equation}
\label{eq:traj}
A(\vec{n}_\alpha)=\langle n_f^\alpha|e^{-i\hat{H}\delta t}|n_N^\alpha\rangle\cdots\langle n_2^\alpha|e^{-i\hat{H}\delta t}|n_1^\alpha\rangle.
\end{equation}
Here  the $n_j^\alpha$ denote the states inserted at the $j$-th time-slice either on the left ($\alpha=L$) or on the right side of the density matrix ($\alpha=R$) in Eq.~\eqref{eq:momgen}. The quantity $A(\vec{n}_\alpha)$ gives the amplitude for a trajectory through Hilbert space, defined by the vector $\vec{n}_\alpha=(n_1^\alpha,\cdots,n_N^\alpha,n_f^\alpha)$. The time integral of the observable $\hat{n}$ over such a discrete trajectory is given by $m_\alpha =\sum_{j}n_j^\alpha\delta t$. Examples of such trajectories are illustrated in Fig.~\ref{fig:trajscem}\,(a) and (b).  Finally, as we are interested in the $\delta t \rightarrow 0$ limit, we neglect terms that are order $(\delta t)^2$ and higher.

Each term in Eq.~\eqref{eq:pofmres} describes the contribution to $P(m)$ from a pair of trajectories $\vec{n}_L$ and $\vec{n}_R$; the second line is the product of probability amplitudes for each of the trajectories, 
weighted by the density matrix element corresponding to the initial ``position" of each trajectory. The trajectories are summed over, given the constraints on the first line. The Kronecker delta enforces the two trajectories to end at the same position and is a consequence of the trace in Eq.~\eqref{eq:momgen}. The Dirac delta tells us that a pair of trajectories contributes to $P(m)$ when $m$ is equal to the \textit{average} of $m_L$ and $m_R$.


While Eq.~\eqref{eq:pofmres} is just a direct representation of the standard FCS $P(m)$ distribution, we immediately notice a rather strange feature:  for a given particular value $m_0$, 
the interference terms between two trajectories can contribute
to $P(m_0)$ even though the corresponding classical probabilities do not. To be explicit, suppose we have a pair of trajectories with $m$ values $m_L$ and $m_R$.  The classical probability from each trajectory, i.e.~the terms proportional to $|A(\vec{n}_L)|^2$ and $|A(\vec{n}_R)|^2$, contribute to $P(m_L)$ and $P(m_R)$ respectively. Their interference terms, i.e.~the terms proportional to $A(\vec{n}_L)A^*(\vec{n}_R)$ and $A(\vec{n}_R)A^*(\vec{n}_L)$, contribute instead to $P(m_L/2+m_R/2)$. If $m_L\neq m_R$, the interference terms are thus separated from their classical probabilities allowing the quasi-probability distribution $P(m)$ to become negative. We thus have one of the key conclusions of our approach:  negativity in the distribution $P(m)$ is directly and necessarily connected to a kind of anomalously strong influence of interferences between pairs of trajectories.

This motivates us to separate the contributions to the sum in Eq.~\eqref{eq:pofmres} into two generic kinds. Terms with $m_L=m_R$ are denoted ``classical" contributions. These yield a total contribution to $P(m)$ which is positive definite. Terms with $m_L\neq m_R$ are denoted ``interference" contributions. These are the interference terms which are separated from their classical probabilities and responsible for any negativities in the FCS. Examples of pairs of trajectories yielding ``classical" and ``interference" contributions to $P(m)$ are illustrated in Fig.~\ref{fig:trajscem}\, (a), (b).

\begin{figure*}[t!]
\centering
\includegraphics[width=\textwidth]{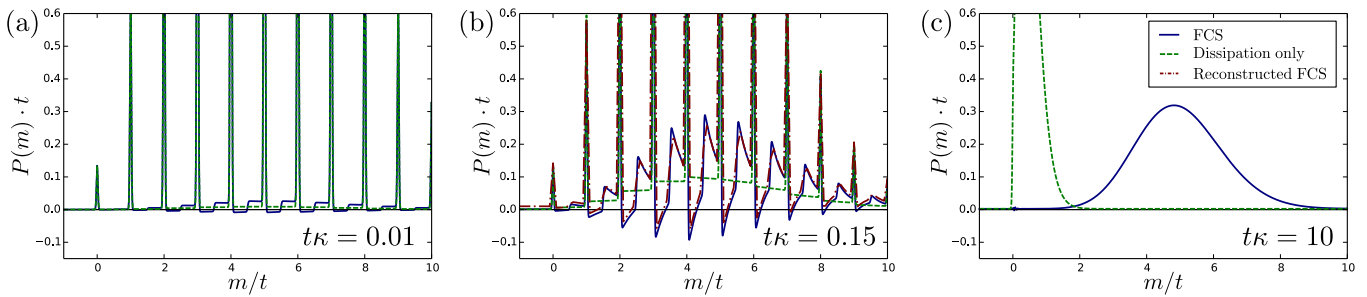}
\caption{Time-evolution of the integrated-energy FCS for a damped cavity initially prepared in a coherent state. The blue curve shows the FCS in the presence of a coherent drive with strength $f/\kappa=\sqrt{5}/2$, where $\kappa$ is the energy damping rate. The green curve shows the FCS describing the dissipative emptying of the cavity ($f=0$). The resulting distribution is fully positive and can be described with a classical model. 
(a) At very short times, the FCS is dominated by peaks at integer $m/t$ reflecting the initial photon distribution. (b) The exchange of photons with the coherent drive and the dissipative bath leads to features in between the peaks which can be understood in terms of the few-jump trajectories. An experimental reconstruction of the FCS (red, dashed-dotted) using an auxiliary qubit detector is feasible using $500$ measurements covering a range of $\lambda$ up to $\lambda_{max}=418.9\kappa$ (see main text). (c) At long times, the FCS is a continuous function peaked around the mean photon number in the cavity. Except for the reconstructed FCS, all distributions are convolved with a sharply peaked Gaussian (width $\sigma=t\kappa/10$) to resolve the Dirac deltas. For all panels, the drive is on resonance $\Delta=0$.}
  \label{fig:time_evolution}
\end{figure*}

\textit{Driven cavity} -- We now illustrate our trajectory approach to FCS by considering a coherently driven bosonic single-mode resonator, first in the absence of any dissipation. This constitutes a simple system which is amenable to cQED \cite{paik:2011,wallraff:2004,kirchmair:2013} and optomechanical \cite{murch:2008,groblacher:2009} experiments. In the frame rotating at the driving frequency, the Hamiltonian of the system reads
\begin{equation}
\label{eq:hamiltonian}
\hat{H}=\Delta \hat{a}^\dag \hat{a} - f(\hat{a}^\dag + \hat{a}),
\end{equation}
where $\Delta$ denotes the detuning of the drive and $f$ the drive strength (which we take to be real without loss of generality). We are interested in the photon number fluctuations, $\hat{n}=\hat{a}^\dagger\hat{a}$. Despite the seemingly trivial nature of the system and its linear dynamics, we are measuring a nonlinear observable, and the integrated energy fluctuations are described by negative FCS \cite{clerk:2011}. 

The coherent drive can induce jumps in the trajectories (i.e.~from one Fock state to another) whereas the detuning introduces a phase factor in $A(\vec{n}_L)A^*(\vec{n}_R)$ whenever $m_L\neq m_R$. For $ft\ll1$, only pairs of trajectories with a low number in jumps will contribute to the FCS and we can make some analytical progress. To this end, we consider an inital Fock state $\hat{\rho}=|n_0\rangle\langle n_0|$ and pairs of trajectories including a total of up to two jumps. The contribution from pairs exhibiting no jumps at all is given by 
\begin{equation}
\label{eq:pofm0}
P_0(m)=\delta(m-n_0t).
\end{equation}
The zero jump contribution thus reflects the initial distribution and does not decay with time. To ensure the normalization of $P(m)$, all contributions with a higher number of jumps must thus average to zero:  this ensures negativities in the FCS as long as the dynamics of the system is non-trivial. These considerations remain valid for an arbitrary initial state.

For an initial Fock state, there is no contribution from pairs of trajectories exhibiting a single jump in total because of the Kronecker delta in Eq.~\eqref{eq:pofmres}. The two jump contribution is discussed in the SM \cite{supplementarx} and plotted in Fig.~\ref{fig:trajscem}\,(c) together with a Monte Carlo simulation of the FCS. As illustrated in Fig.~\ref{fig:trajscem}\,(c), the distribution $P(m)$ shows a highly non-trivial behavior and becomes negative over a substantial range of its argument. The jump at $m/t=n_0$ as well as the kinks at $m/t=n_0\pm1/2$ are a consequence of the discreteness of photon numbers and can be well understood in terms of the few trajectories that contribute at short times (see SM \cite{supplementarx}).

We stress that these unusual short-time features also occur for different choices of initial states, including a coherent state; in that case, our calculations agree with the approach used in Refs.~\cite{clerk:2007, clerk:2011} (see also Fig.~\ref{fig:time_evolution}).  The presence of negative FCS is thus not a function of the initial state, but rather reflects the non-classicality of the system dynamics; this is in stark contrast to the Wigner function (where coherent states exhibit no negativity). We thus conclude that even systems which remain in a seemingly near-classical state at all times can exhibit extremely non-classical behavior in their dynamics.

\textit{Additional degrees of freedom} -- 
Equation \ref{eq:pofmres} (and the single resonator example) discussed so far are somewhat special cases,  in that the relevant dynamics only involves a single degree of freedom.  As we now show, if the dynamics starts to couple to additional degrees of freedom, negativity can be rapidly lost, as there is a strong suppression of the required ``interference" contributions.


Consider first the situation where the additional degrees of freedom correspond to a dissipative environment; for concreteness, we return to our example of a driven resonator, and add a coupling to a Markovian bath.  In such a situation, the contribution of the bath to the dynamics can be modelled in terms of dissipative quantum jumps, in complete analogy to how they are treated in the standard quantum trajectory approach of quantum optics \cite{carmichael:book}. These dissipation-induced jumps are described by the superoperators
\begin{equation}
\label{eq:dissjumps}
\mathcal{J}_\downarrow\hat{\rho}=\kappa(n_B+1) \hat{a}\hat{\rho} \hat{a}^\dagger,\hspace{1cm}\mathcal{J}_\uparrow\hat{\rho}=\kappa n_B \hat{a}^\dag\hat{\rho} \hat{a},
\end{equation}
where $\kappa$ is the energy damping rate and $n_B$ is the thermal occupation number of the bath at the cavity frequency. The first (second) term describes photons which are lost to (gained from) the bath. 

We can again incorporate these jump operators into a path-integral expression for the FCS distribution function $P(m)$, see SM \cite{supplementarx}.  Similar to standard quantum trajectory theory, the dissipation correlates the behavior of the left
and right trajectories, thus suppressing the negativity-induced  ``interference" contributions (which require distinct trajectories on the left and on the right).  
For a purely dissipative process, the left and the right trajectories are always identical and the FCS always positive, being a simple sum of classical probabilities. In this case, the FCS recovers the results obtained by classical master equations (see Fig.~\ref{fig:time_evolution}). Details on the dissipative FCS calculation are provided in the SM \cite{supplementarx}, as well as a discussion on how coupling coherently to an additional degree of freedom also suppresses negativity.

\textit{Time evolution of the FCS} -- To stress the utility of our approach we consider the time evolution of the FCS in an experimentally relevant system. To this end, we add dissipation to our driven cavity system resulting in the Lindblad master equation
\begin{equation}
\label{eq:lindblad}
\frac{d\hat{\rho}}{dt}=-i[\hat{H},\hat{\rho}]-\frac{\kappa}{2}\left\{\hat{a}^\dag \hat{a},\hat{\rho} \right\}+\kappa \hat{a}\hat{\rho} \hat{a}^\dagger,
\end{equation}
where $\hat{H}$ is given in Eq.~\eqref{eq:hamiltonian}. As an initial state, we take the steady state solution which is given by a coherent state with an average photon number $n_D=4f^2/(\kappa^2+4\Delta^2)$. Since in this case the Wigner function is Gaussian at all times, we use the method described in detail in Refs.~\cite{clerk:2007,clerk:2011} to calculate the moment generating function. The time evolution of the resulting FCS is illustrated in Fig.~\ref{fig:time_evolution}. At very short times [cf.~panel (a)], the FCS is dominated by sharp peaks at integer $m/t$ corresponding to trajectories where the photon number remains constant. At large times [cf.~panel (c)], the FCS is a smooth function centered around the mean photon number $n_D$. At times where the trajectories with few jumps dominate [cf.~panel (b)], the FCS exhibits features in between the peaks at integer $m/t$. For a purely dissipative process, the FCS is continuous in between the peaks and can be captured by a classical calculation involving only occupation probabilities. 
In the presence of a coherent drive, the FCS exhibits a surprising shape with discontinuities at half integer $m/t$. In complete analogy to Fig.~\ref{fig:trajscem}\,(c), this can be well understood in terms of the few-jump trajectories and ultimately results from the discreteness of the number of photons. The jumps at half-integer $m/t$ are a consequence of the coherences in the initial state. Our approach thus allows for a quantitative understanding of the non-trivial short-time FCS.

\textit{Reconstructing the FCS} -- As discussed in detail in the SM \cite{supplementarx}, measurement noise (uncertainty and backaction) will often mask the sharp features which are characteristic for the short-time regime. Motivated by the exceptional quality of cQED experiments \cite{wallraff:2004,paik:2011,kirchmair:2013}, we thus dedicate the remainder of this letter to the reconstruction of the FCS by coupling a qubit dispersively to the observable of interest. As discussed above, the (unperturbed) moment generating function \eqref{eq:momgen} can be accessed through the off-diagonal density matrix element of a qubit which couples to the observable of interest with the coupling Hamiltonian $\hat{H}_c=\lambda\hat{n}\hat{\sigma}_z/2$. Since the FCS is given by the Fourier transform of the moment generating function, the latter would have to be measured for all possible values of the coupling strength $\lambda$ in order to faithfully reconstruct the FCS. Here we are interested in how well this reconstruction performs if the measurements are limited in number and the coupling strength can not exceed a maximal value.

As shown in Fig.~\ref{fig:time_evolution}\,(b), a maximal coupling strength of $\lambda_{max}\approx 420\kappa$ with $500$ equally spaced measurement points is sufficient to reconstruct most features of the FCS. As discussed in the SM \cite{supplementarx}, this procedure is robust against uncertainties in the coupling strength up to a magnitude of $\sim\kappa/2$. However, some care has to be taken in the choice of $\lambda_{max}$ and the post processing of the measured values.

\textit{Conclusions} -- By unraveling the FCS in terms of trajectories through Hilbert space, we demonstrated that negative FCS arise from a peculiar interference effect, where the interference contribution from a pair of trajectories can contribute without the corresponding classical probabilities.  Our approach highlights how negative FCS are directly tied to non-classical
\textit{dynamics}, in contrast to standard quasi-probabilities which characterize non-classical \textit{states}.  We hope that the understanding of negative FCS presented here will inspire further work on non-classical dynamical processes, as well on experiments to measure these effects.

\textit{Acknowledgements} -- We acknowledge useful comments from D. Dasenbrook, Y. Nazarov, P. Solinas, and S. Gasparinetti. This work was supported by NSERC. P. P. H. acknowledges funding from the Swiss NSF.

\bibliography{biblio}

\clearpage
\widetext
\begin{center}
\textbf{\large Supplement: Negative Full Counting Statistics Arise From Interference Effects}
\end{center}
\setcounter{equation}{0}
\setcounter{figure}{0}
\setcounter{table}{0}
\setcounter{page}{1}
\makeatletter
\renewcommand{\theequation}{S\arabic{equation}}
\renewcommand{\thefigure}{S\arabic{figure}}

\section{I. Additional degrees of freedom}
In this section, we show how additional degrees of freedom which couple to the operator of interest suppress the interference terms and thus reduce the negativity in the FCS.
\subsection{A. One additional bosonic mode}
We first investigate the effect of one additional bosonic mode (described by the operator $\hat{a}_\phi$) on the FCS of the counted bosonic mode (described by the operator $\hat{a}_n$). We split the Hamiltonian in three parts,
\begin{equation}
\hat{H}=\hat{H}_n+\hat{H}_\phi+\hat{H}_{n\phi},
\end{equation}
where the first two terms only act on the counted and the additional mode respectively. Similarly to the derivation of Eq.~\eqref{eq:pofmres}, we now insert identities into the definition of the moment generating function [cf.~Eq.~\eqref{eq:momgen}]. Unlike the counted mode, we resolve the additional mode in terms coherent states
\begin{equation}
\label{eq:idcountadd}
\mathbb{I}=\sum\limits_{n}|n\rangle\langle n|\otimes\int d\bar{\phi} d\phi |\phi\rangle\langle \phi|,\hspace{2cm}|\phi\rangle=e^{-\frac{|\phi|^2}{2}}\sum\limits_{n=0}^{\infty}\frac{\phi^n}{\sqrt{n!}}|n_\phi\rangle.
\end{equation}
We find that Eq.~\eqref{eq:pofmres} has to be replaced with a similar expression which includes an average over the additional mode
\begin{equation}
\label{eq:pofmresphi}
P_\phi(m)=\sum_{\vec{n}_L,\vec{n}_R}\delta_{n_f^L,n_f^R}\delta\left(m-\frac{1}{2}m_L-\frac{1}{2}m_R\right)\Big\langle\delta(\phi_f^L-\phi_f^R)\delta(\bar{\phi}_f^L-\bar{\phi}_f^R)\langle n_1^L|\hat{\rho}(\bar{\phi}_1^L,\phi_1^R)|n_1^R\rangle A(\vec{n}_L,\vec{\phi}_L)A^*(\vec{n}_R,\vec{\phi}_R)\Big\rangle_\phi,
\end{equation}
with the amplitudes
\begin{equation}
\label{eq:trajphi}
A(\vec{n}_\alpha,\vec{\phi}_\alpha)=\langle n_f^\alpha|e^{-i\hat{H}_{n}\delta t} e^{-i\hat{H}_{n\phi}(\bar{\phi}_f^\alpha,\phi_N^\alpha)\delta t}|n_N^\alpha\rangle\cdots \langle n_2^\alpha|e^{-i\hat{H}_{n}\delta t} e^{-i\hat{H}_{n\phi}(\bar{\phi}_2^\alpha,\phi_1^\alpha)\delta t}|n_1^\alpha\rangle.
\end{equation}
Here $O(\bar{\phi},\phi')=\langle \phi|\hat{O}|\phi'\rangle$
and the average over the fields reads
\begin{equation}
\Big\langle(\cdots)\Big\rangle_\phi=\int\prod_{j=1}^{2N+2} d \bar{\phi}_jd\phi_je^{i\mathcal{G}[\bar{\phi},\phi]}(\cdots).
\end{equation}
The index $j$ is an index which follows the Keldysh contour
\begin{equation}
\phi_1=\phi_1^L,\hspace{.2cm}\cdots,\hspace{.2cm}\phi_N=\phi_N^L,\hspace{.2cm}\phi_{N+1}=\phi_f^L,\hspace{.2cm}\phi_{N+2}=\phi_N^R,\hspace{.2cm}\cdots,\hspace{.2cm}\phi_{2N+1}=\phi_1^R,\hspace{.2cm}\phi_{2N+2}=\phi_f^R,
\end{equation}
and the (discrete) Greens function for the additional cavity fields reads
\begin{equation}
\mathcal{G}[\bar{\phi},\phi]=\sum\limits_{j=2}^{2N+1}\delta t_j\left[i\bar{\phi}_j\frac{\phi_j-\phi_{j-1}}{\delta t_j}-H_\phi(\bar{\phi}_j,\phi_{j-1})\right]+i|\phi_1|^2,
\end{equation}
where $\delta t_{j\leq N+1}=\delta t$ and $\delta t_{j> N+1}=-\delta t$.

To illustrate the effect of the additional mode, we discuss two examples for the coupling Hamiltonian. For simplicity, we assume that the additional cavity has no internal dynamics and is initially in a coherent state
\begin{equation}
\hat{H}_\phi=0,\hspace{2cm}\hat{\rho}=\hat{\rho}_n\otimes|\phi_0\rangle\langle\phi_0|.
\end{equation}

For the optomechanical coupling Hamiltonian
\begin{equation}
\hat{H}_{n\phi}=\chi\hat{n}\left(\hat{a}_\phi^\dagger+\hat{a}_\phi\right),
\end{equation}
we find
\begin{equation}
P_\phi(m)=\sum_{\vec{n}_L,\vec{n}_R}\delta_{n_f^L,n_f^R}\delta\left(m-\frac{1}{2}m_L-\frac{1}{2}m_R\right)e^{-i\chi(\bar{\phi}_0+\phi_0)(m_L-m_R)}e^{-\frac{\chi^2}{2}(m_L-m_R)^2}\langle n_1^L|\hat{\rho}|n_1^R\rangle A(\vec{n}_L)A^*(\vec{n}_R),
\end{equation}
where $A(\vec{n}_\alpha)$ is given in Eq.~\eqref{eq:traj}.
The first exponential is the average detuning induced by the additional bosonic mode while the second exponential decreases the weight of the non-classical trajectories due to the uncertainty in the induced detuning. This reduction is analogous to the backaction of a heavy mass detector [cf.~Eq.~\eqref{eq:fcshm}]. In both cases, the detuning is described by a Gaussian random variable which remains constant in time due to the lack of dynamics of the additional degree of freedom coupling to $\hat{n}$.

For the coupling Hamiltonian (i.e.~a cross-Kerr interaction)
\begin{equation}
\hat{H}_{n\phi}=\chi\hat{n}\hat{a}_\phi^\dagger \hat{a}_\phi,
\end{equation}
we find
\begin{equation}
\begin{aligned}
P_\phi(m)=&\sum_{\vec{n}_L,\vec{n}_R}\delta_{n_f^L,n_f^R}\delta\left(m-\frac{1}{2}m_L-\frac{1}{2}m_R\right)\exp\left[|\phi_0|^2\left(e^{-i\chi(m_L-m_R)}-1\right)\right]\langle n_1^L|\hat{\rho}|n_1^R\rangle A(\vec{n}_L)A^*(\vec{n}_R)\\
=&\sum_{\vec{n}_L,\vec{n}_R}\delta_{n_f^L,n_f^R}\delta\left(m-\frac{1}{2}m_L-\frac{1}{2}m_R\right)e^{-|\phi_0|^2}\sum\limits_{p=0}^{\infty}\frac{|\phi_0|^{2p}}{p!}e^{-ip\chi(m_L-m_R)}\langle n_1^L|\hat{\rho}|n_1^R\rangle A(\vec{n}_L)A^*(\vec{n}_R).
\end{aligned}
\end{equation}
Again, the additional mode has the effect of an induced, random detuning. Here the different number states which contribute to the coherent state $|\phi_0\rangle$ lead to different detuning strengths.

Analogous to the above discussion, we would expect a coupling Hamiltonian of the form $\hat{H}_{n\phi}=\chi(\hat{a}^\dagger_n \hat{a}_\phi+\hat{a}^\dagger_\phi \hat{a}_n)$ to have the same effect as a fluctuating driving force. Additional degrees of freedom can thus introduce uncertainties in different parameters. The distribution of these parameters depends on the initial state and the dynamics of the additional degrees of freedom.

\subsection{B. FCS for a time evolution governed by a Lindblad master equation}
We now consider the FCS in the presence of a Markovian bath. The time evolution is governed by the master equation
\begin{equation}
\label{eq:master}
\frac{d\hat{\rho}}{dt}=-i[\hat{H},\hat{\rho}]-\kappa\left(n_B+1/2\right)\left\{\hat{a}^\dag \hat{a},\hat{\rho} \right\}-\kappa n_B\hat{\rho}+\kappa\left(n_B+1\right) \hat{a}\hat{\rho} \hat{a}^\dagger+\kappa n_B \hat{a}^\dagger\hat{\rho} \hat{a}=\mathcal{L}\hat{\rho},
\end{equation}
where $n_B$ denotes the occupation number of the bath at the cavity frequency. We now split the superoperator $\mathcal{L}$ into two parts: a Hamiltonian evolution with a non-hermitian Hamiltonian and a dissipative jump superoperator $\mathcal{L}=\mathcal{L}_\mathcal{H}+\mathcal{J}$. These superoperators act on the density matrix as
\begin{equation}
\begin{aligned}
&\mathcal{L}_\mathcal{H}\hat{\rho}=-i[\hat{H},\hat{\rho}]-\kappa\left( n_B+1/2\right)\left\{\hat{a}^\dag \hat{a},\hat{\rho} \right\}-\kappa n_B \hat{\rho}\hspace{1cm}\Leftrightarrow\hspace{1cm}e^{\mathcal{L}_\mathcal{H}t}\hat{\rho}=e^{-i\mathcal{\hat{H}}t}\hat{\rho} e^{i\mathcal{\hat{H}}^\dag t},\\
&\mathcal{J}\hat{\rho}=\kappa\left(n_B+1\right)\kappa \hat{a}\hat{\rho} \hat{a}^\dagger+\kappa n_B \hat{a}^\dagger\hat{\rho} \hat{a},
\end{aligned}
\end{equation}
with the non-hermitian Hamiltonian
\begin{equation}
\label{eq:nonhermham}
\mathcal{\hat{H}}=\hat{H}-i\kappa\left(n_B+1/2\right) \hat{a}^\dag \hat{a}-i\frac{\kappa}{2} n_B.
\end{equation}
We can then unravel the evolution of the master equation times $t_i$ \cite{carmichael:book}
\begin{equation}
\label{eq:unravel}
\hat{\rho}(t)=e^{\mathcal{L}t}\hat{\rho}(0)=\sum\limits_{p=0}^{\infty}\int\limits_{0}^{t}dt_p\int\limits_{0}^{t_p}dt_{p-1}\cdots\int\limits_{0}^{t_2}dt_1e^{\mathcal{L}_\mathcal{H}(t-t_p)}\mathcal{J}e^{\mathcal{L}_\mathcal{H}(t_p-t_{p-1})}\cdots\mathcal{J}e^{\mathcal{L}_\mathcal{H}t_1}\hat{\rho}(0).
\end{equation}
This corresponds to a Hamiltonian evolution with a non-unitary Hamiltonian which is interrupted by $p$ dissipative quantum jumps occurring at times $t_j$ which consist of a photon entering or leaving the cavity. We can express the Hamiltonian evolution in terms of trajectories as we did in the derivation of Eq.~\eqref{eq:pofmres}
\begin{equation}
\label{eq:superham}
e^{\mathcal{L}_\mathcal{H}t}\hat{\rho}=\sum_{\vec{n}_L,\vec{n}_R}e^{-\kappa\left(n_B+\frac{1}{2}\right)\left(m_L+m_R\right)}e^{-\kappa n_B t}\langle n_1^L|\hat{\rho}|n_1^R\rangle A(\vec{n}_L)A^*(\vec{n}_R)|n_f^L\rangle \langle n_f^R|,
\end{equation}
where again $m_\alpha=\sum\limits_{j=1}^{N}n_j^\alpha \delta t$. The dissipation thus introduces dissipative jumps which happen on the left and on the right trajectory simultaneously and it reduces the weight of terms with a high $m$.

\begin{figure*}[t!]
\centering
\includegraphics[width=.8\textwidth]{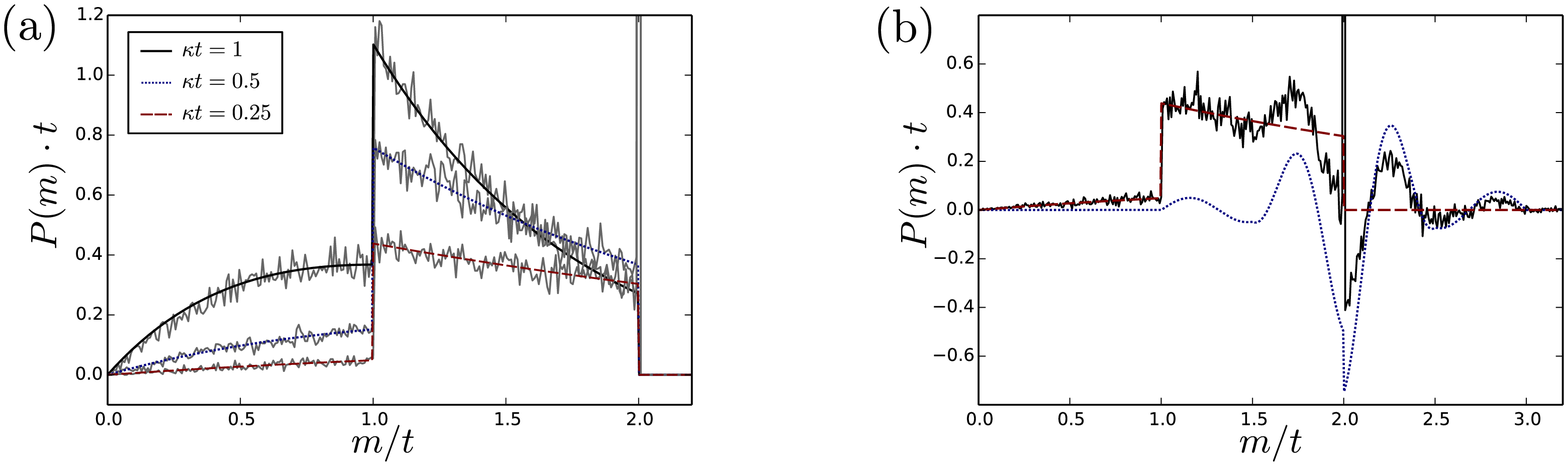}
\caption{Effect of dissipation on the FCS for an initial Fock state $n_0=2$.(a) Purely dissipative process ($f=0$) for different damping parameters $\kappa$. For small $\kappa$ only one photon leaks out of the cavity. As $\kappa$ increases, the probability of both photons leaking out increases. (b) Combined effect of coherent drive and dissipation for $\kappa t=0.25$ $ft=0.25$, $\Delta t=5$. As a comparison, the FCS for $\kappa=0$ is plotted in blue (dotted) and the FCS for $f=0$ is plotted in red (dashed). Noisy lines are calculated using a Monte Carlo simulation with $50,000$ trajectories, steady lines are analytical results (omitting singular contributions).}
  \label{fig:dissipation}
\end{figure*}

As discussed in the main text, the moment generating function is given by the off-diagonal element of the reduced density matrix of a qubit that couples linearly to the operator $\hat{n}$. Replacing $\hat{H}$ with $\hat{H}\pm\lambda \hat{n}/2$ on the left/right trajectory, we can derive a quasi-probability analogously to Eq.~\eqref{eq:pofmres}
\begin{equation}
P(m)=\sum_{\vec{n}_L,\vec{n}_R}\delta_{n_f^L,n_f^R}\delta\left(m-\frac{1}{2}m_L-\frac{1}{2}m_R\right)e^{-\kappa\left(n_B+\frac{1}{2}\right)\left(m_L+m_R\right)}e^{-\kappa n_B t}\mathcal{T}_\kappa(\vec{n}_L,\vec{n}_R),
\end{equation}
with
\begin{equation}
\mathcal{T}_\kappa(\vec{n}_L,\vec{n}_R)=\sum\limits_{p=0}^{\infty}\int\limits_{0}^{t}dt_p\int\limits_{0}^{t_p}dt_{p-1}\cdots\int\limits_{0}^{t_2}dt_1\langle n_f^L|e^{\mathcal{P}(\vec{n}_L,\vec{n}_R;t,t_p)}\mathcal{J}e^{\mathcal{P}(\vec{n}_L,\vec{n}_R;t_p,t_{p-1})}\cdots\mathcal{J}e^{\mathcal{P}(\vec{n}_L,\vec{n}_R;t_1,0)}\hat{\rho}(0)|n_f^R\rangle.
\end{equation}
Here the superoperator appearing between the dissipative jumps evolves the density matrix along a given pair of trajectories determined by the elements of the vectors $\vec{n}_\alpha$ which correspond to times between the two time arguments (where $t_{p+1}=t$ and $t_0=0$). For $l<k$, we have
\begin{equation}
\begin{aligned}
e^{\mathcal{P}(\vec{n}_L,\vec{n}_R;k\delta t,l\delta t)}\hat{\rho}=&|n^L_{k+1}\rangle\langle n^L_{k+1}|e^{-i\hat{H}\delta t}|n^L_{k}\rangle\cdots\langle n^L_{l+2}| e^{-i\hat{H}\delta t}|n_{l+1}^L\rangle\\\times&\langle n_{l+1}^L|\hat{\rho}|n_{l+1}^R\rangle\langle n_{l+1}^R|e^{i\hat{H}\delta t}|n^R_{l+2}\rangle\cdots\langle n^R_{k}|e^{i\hat{H}\delta t}|n^R_{k+1}\rangle\langle n^R_{k+1}|,
\end{aligned}
\end{equation}
and
\begin{equation}
e^{\mathcal{P}(\vec{n}_L,\vec{n}_R;t,0)}\hat{\rho}=\langle n_1^L|\hat{\rho}|n_1^R\rangle A(\vec{n}_L)A^*(\vec{n}_R)|n_f^L\rangle \langle n_f^R|,
\end{equation}
The effect of dissipation is illustrated in Fig.~\ref{fig:dissipation}.

\section{II. Added noise of the measurement}
As mentioned in the introduction, the FCS distribution $P(m)$ represents the intrinsic fluctuations of the system.  It can be used to predict the distribution of outcomes in an actual measurement setup, but one must convolve-in the effects of the added noise of the measurement apparatus (both imprecision noise and backaction noise).  While this has been discussed in the past, our path-integral approach gives a particularly intuitive way to understand measurement-noise effects.
We consider the generic heavy-mass detector used in Ref.~\cite{nazarov:2003}. In this setup, the observable to be measured $\hat{n}$ is linearly coupled to the position $\hat{x}$ of an infinite mass via the coupling Hamiltonian $H_{c}=-A\hat{x}\hat{n}$. The interaction between the system and the detector is then turned on for a time $t$ and the FCS can be inferred from the momentum distribution of the detector after the measurement. The initial momentum uncertainty $\sigma_p$ of the detector gives rise to measurement imprecision, while the uncertainty in its position $\sigma_x$ induces backaction. The measurement imprecision has the effect of convolving the FCS with a Gaussian of width $\sigma_{imp}=\sigma_p/A$. The backaction can be incorporated by introducing an additional term $H_{BA}=\Delta'\hat{n}$ in the Hamiltonian, where $\Delta'$ is a random variable with a Gaussian distribution of width $\sigma_{BA}=A\sigma_x$ \cite{clerk:2011}. Note that although random, $\Delta'$ is constant in time because the detector remains at a fixed position (due to its infinite mass). The measured probability distribution reads
\begin{equation}
\label{eq:fcshm}
P_{d}(m)=\frac{1}{\sqrt{2\pi}\sigma_{imp}}\sum_{\vec{n}_L,\vec{n}_R}\delta_{n_f^L,n_f^R}e^{-\frac{\left(m-\frac{1}{2}m_L-\frac{1}{2}m_R\right)^2}{2\sigma^2_{imp}}} e^{-\frac{\sigma_{BA}^2}{2}\left(m_L-m_R\right)^2}\langle n_1^L|\hat{\rho}|n_1^R\rangle A(\vec{n}_L)A^*(\vec{n}_R),
\end{equation}
where the subscript $d$ reminds us of the presence of a detector. Measurement imprecision replaces the Dirac delta in Eq.~\eqref{eq:pofmres} with a Gaussian, smearing out any sharp features in the FCS. The backaction exponentially reduces interference contributions with $m_L\neq m_R$. The Heisenberg uncertainty principle applied to the detector implies the relation $\sigma_{BA}\sigma_{imp}\geq 1/2$.

In the case of a quantum limited measurement, where $\sigma_{BA}\sigma_{imp}=1/2$, the probability distribution reduces to
\begin{equation}
\label{eq:fcshmql}
P_{d}(m)=\sqrt{\frac{2}{\pi}}\sigma_{BA}\sum_{\vec{n}_L,\vec{n}_R}\delta_{n_f^L,n_f^R}\langle n_1^L|\hat{\rho}|n_1^R\rangle e^{-(m-m_L)^2\sigma_{BA}^2}e^{-(m-m_R)^2\sigma_{BA}^2}A(\vec{n}_L)A^*(\vec{n}_R).
\end{equation}
Using $\delta_{n_f^L,n_f^R}=\sum_n\langle n_f^R|n\rangle\langle n | n_f^L\rangle$ and $\hat{\rho}=\sum_kp_k|\psi_k\rangle\langle\psi_k|$, we find
\begin{equation}
P_{d}(m)=\sqrt{\frac{2}{\pi}}\sigma_{BA}\sum\limits_{n,k}p_k\left|\sum_{\vec{n}_L}\langle n_1^L|\psi_k\rangle\langle n|n_f^L\rangle e^{-(m-m_L)^2\sigma_{BA}^2}A(\vec{n}_L)\right|^2,
\end{equation}
which is evidently positive.

\section{III. Two jump contribution to the FCS of an initial Fock state}
Here we explicitly calculate the two jump contribution to the FCS displayed in Fig.~\ref{fig:trajscem}. The term where the two jumps happen on different trajectories can be divided into two terms. One where the jumps increase the photon number ($|n_0\rangle\rightarrow|n_0+1\rangle$) and one where the jumps reduce the photon number ($|n_0\rangle\rightarrow|n_0-1\rangle$). Starting from Eq.~\eqref{eq:pofmres} with the Hamiltonian given in Eq.~\eqref{eq:hamiltonian} and an initial Fock state $\hat{\rho}=|n_0\rangle\langle n_0|$, we find
\begin{equation}
\label{eq:p2barup}
\begin{aligned}
\bar{P}_{2\uparrow}(m)=&\sum\limits_{l,r=0}^{N}\delta\left(m-(n_0+1)t+\frac{1}{2}(l\delta t+r\delta t)\right)\langle n_0+1|e^{-i\Delta \hat{n}(t-l\delta t)}|n_0+1\rangle\langle n_0+1|if\delta t\hat{a}^\dag |n_0\rangle\langle n_0|e^{-i\Delta \hat{n}l\delta t}|n_0\rangle\\&\times\langle n_0|e^{i\Delta\hat{n}r\delta t}|n_0\rangle\langle n_0| (-i)f\delta t\hat{a}|n_0+1\rangle\langle n_0+1|e^{i\Delta\hat{n}(t-r\delta t)}|n+1\rangle\\=&(n_0+1)f^2\sum\limits_{l=0}^{N}\delta t\sum\limits_{r=0}^{N}\delta t\delta\left(m-(n_0+1)t+\frac{1}{2}(l\delta t+r\delta t)\right)e^{i\Delta(l\delta t-r\delta t)},
\end{aligned}
\end{equation}
where the bar denotes that the two jumps happen on different trajectories and the subscript denotes that the term includes two jumps which increase the photon number by one. The sum is over all possible times at which the jumps may occur. In the continuum limit, we find
\begin{equation}
\bar{P}_{2\uparrow}(m)=(n_0+1)f^2\int\limits_{0}^{t}dt_L\int\limits_{0}^{t}dt_R\delta\left(m-(n_0+1)t+\frac{1}{2}(t_L+t_R)\right)e^{i\Delta(t_L-t_R)}.
\end{equation}
Evaluating the integral and including the term where the jumps decrease the photon number, we find the contribution where the two jumps happen on different trajectories
\begin{equation}
\label{eq:trajp2a}
\bar{P}_2(m)=\frac{2f^2}{\Delta}\left\lfloor \frac{m}{t}+1\right\rfloor\sin\left(2\Delta \left|m-\left\lfloor \frac{m}{t}+\frac{1}{2}\right\rfloor t\right|\right) \hspace{1cm}\text{for }m\in[(n_0-1)t,(n_0+1)t],
\end{equation}
and zero otherwise. Here $\lfloor\cdot\rfloor$ denotes the floor function. There are kinks whenever $m/t$ is a multiple of $1/2$. This is a consequence of the fact that a \textit{single} trajectory with one upward/downward jump contributes to $m$ with $m_\alpha/2 \in [n_0/2, n_0/2\pm 1/2]$.

A similar calculation yields the contribution where the two jumps happen on the same trajectory
\begin{equation}
\label{eq:trajp2b}
\tilde{P}_2(m)=-8f^2\left\lfloor \frac{m}{t}+1\right\rfloor\left|m+\frac{1}{2}t-\left\lfloor \frac{m}{t}+1\right\rfloor t\right|\cos\left[2\Delta(m-n_0t)\right]\hspace{1cm}\text{for }m\in[(n_0-1/2)t,(n_0+1/2)t],
\end{equation}
and zero otherwise. There is a jump at $m/t=n_0$ reflecting the fact that for higher (lower) $m$ the photon number is raised (lowered) in between the two jumps.
Since one of the trajectories does not exhibit any jumps, the last term is non-zero only if $|m-n_0t|<1/2$. The above equations are plotted in Fig.~\ref{fig:trajscem}\,(c).

\section{IV. Reconstructing the FCS: postprocessing the data}
\begin{figure*}[h!]
\centering
\includegraphics[width=\textwidth]{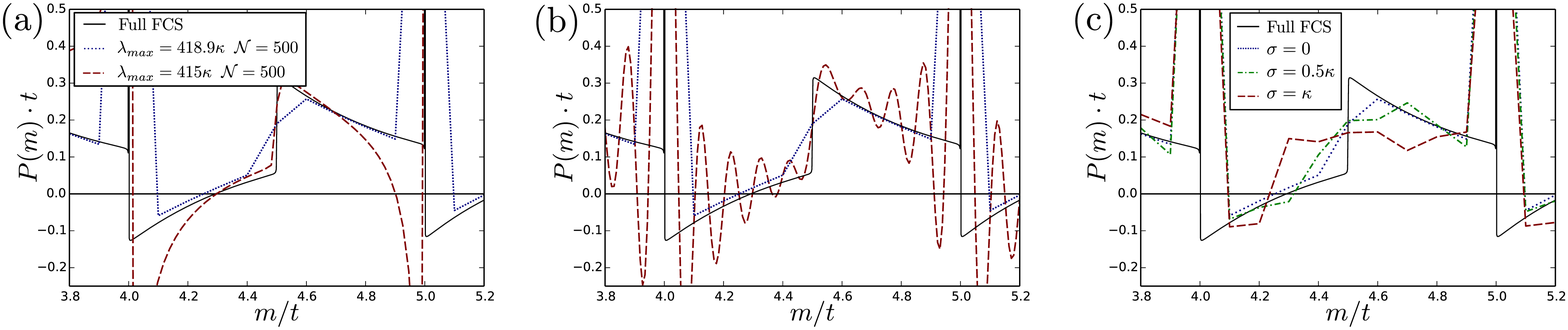}
\caption{Possible issues when reconstructing the FCS. All panels correspond to a zoom in of Fig.~\ref{fig:time_evolution}\,(b) in the main text with the full FCS (without the convolution with a Gaussian) given by the black (solid) line and the reconstructed FCS given by the blue (dotted) line. (a) Sensitivity of the reconstructed FCS to the choice of $\lambda_{max}$. (b) Red (dashed) line is obtained by evaluating Eq.~\eqref{eq:papprox} at all values of $m$. (c) Reconstruction of the FCS if the coupling strengths are normally distributed with width $\sigma$. In terms of the energy damping rate $\kappa$, the observation time is $t\kappa=0.15$.}
  \label{fig:lammax}
\end{figure*}

To reconstruct the FCS with only a limited number of measurements $\mathcal{N}$ up to a maximal coupling strength $\lambda_{max}$, we approximate the Fourier transform in Eq.~\eqref{eq:momgen} by
\begin{equation}
\label{eq:papprox}
P(m)\approx\frac{\lambda_{max}}{\mathcal{N}\pi}\Re\left\{\sum\limits_{n=0}^{\mathcal{N}-1}\Lambda\left(\lambda=\frac{n\lambda_{max}}{\mathcal{N}}\right)e^{2\pi i\frac{n\tilde{m}}{\mathcal{N}}}\right\},
\end{equation}
where we made use of the relation $\Lambda(\lambda)=\Lambda^*(-\lambda)$, which guarantees the FCS to be real, and introduced $\tilde{m}=\frac{\lambda_{max}}{2\pi}m$.
The discontinuous features at multiples of half-integer values for $m/t$ suggest that the moment generating function has terms that oscillate with $\lambda t/2$ which decay very slowly. To properly take into account these oscillations, we need to make sure that our window of integration (here summation) includes an integer number of oscillations. We therefore choose $\frac{\lambda_{max}}{2}t=n2\pi$
with $n\in\mathbb{N}$. That this condition is indeed important is illustrated in Figure~\ref{fig:lammax}\,(a).

In Eq.~\eqref{eq:papprox}, only integer values of $\tilde{m}$ have been used to create Fig.~\ref{fig:time_evolution}\,(b) which allows us to use the fast Fourier transform. Naively, one might expect that an additional evaluation of Eq.~\eqref{eq:papprox} for non-integer $\tilde{m}$ would lead to a better reconstruction of the FCS without the need of more measurement points. However, similarly to the above condition for $\lambda_{max}$, the summand in Eq.~\eqref{eq:papprox} would then include frequencies which fit a non-integer number of oscillations in the window of summation. This itself leads to an oscillatory behavior as illustrated in Fig.~\ref{fig:lammax}\,(b).

These observations might give the impression that a high degree of control over the coupling strength is necessary and that a small deviation from the desired value could make a faithful reconstruction of the FCS impossible. Fortunately, this is not the case. This precision is only required for the post processing of the data. In Fig.~\ref{fig:lammax}\,(c), we reconstruct the FCS using coupling strengths that are normally distributed around their desired values with a width $\sigma$. A value of $\sigma=\kappa/2$ still yields a good reconstruction of the FCS.

It therefore seems achievable to reconstruct the FCS by a finite number of feasible measurements. In order to reduce $\lambda_{max}$, $\kappa$ should be chosen as small as possible. However, if it becomes to small, the uncertainty in the actual coupling strength can complicate the reconstruction.

\end{document}